\documentclass[onecolumn,showpacs,amsmath,amssymb,prd,nofootinbib]{revtex4-2}
\usepackage{graphicx}
\usepackage{epsfig}
\usepackage{enumerate}

\def\beq{\begin{equation}}
\def\eeqno#1{\label{#1}\end{equation}}

\def\rarrow{\rightarrow }

\def\dleft{\rlap{{\it D}}\raise 8pt
\hbox{$\scriptscriptstyle\Leftarrow$}}
\def\dright{\rlap{{\it
D}}\raise 8pt\hbox{$\scriptscriptstyle\Rightarrow$}}

\def\az{a_{0}}

\def\l0{\ell_{0}}

\def\rar{\rightarrow}
\def\s{\sigma}

\def\a{\alpha}
\def\b{\beta}
\def\c{\gamma}
\def\l{\lambda}

\def\f{\phi}
\def\t{\theta}

\def\r{\rho}
\def\m{\mu}
\def\n{\nu}

\def\av#1{\langle#1\rangle}

\def\Ao{\mathcal{A}(\o)}
\def\A{\mathcal{A}}

\def\o{\omega}

\def\vp{{\bf p}}

\def\d{\delta}

\def\a{\alpha}
\def\xlimin{{x\rarrow\infty \atop{\raise 1pt\hbox to 30pt
{\rightarrowfill}}}}
\def\limlim#1#2{{#1\rarrow #2 \atop{\raise 1pt\hbox to 30pt
{\rightarrowfill}}}}

\def\vr{{\bf r}}

\def\vR{{\bf R}}
\def\vv{{\bf v}}

\def\vg{{\bf g}}

\def\va{{\bf a}}

\def\vf{{\bf f}}
\def\vF{{\bf F}}
\def\vP{{\bf P}}
\def\vJ{{\bf J}}
\def\hro{\hat {\bf  r}(\o)}
\def\hFo{\hat {\bf  F}(\o)}
\def\hvo{\hat {\bf  v}(\o)}

\def\hao{\hat {\bf  a}(\o)}
\def\hro{\hat {\bf  r}(\o)}
\def\hvo{\hat {\bf  v}(\o)}

\def\haNo{\hat {\bf  a}_N(\o)}
\def\hFo{\hat {\bf F}(\o)}

\def\grad{\vec\nabla}
\def\div{\vec \nabla\cdot}
\def\gf{\grad\phi}

\def\I{\mathcal{I}}
\def\J{\mathcal{J}}
\def\hao{\hat {\bf  a}(\o)}

\def\haNo{\hat {\bf  a}_N(\o)}
\def\I{\mathcal{I}}

\def\gN{g\_N}
\def\Ao{\mathcal{A}(\o)}

\def\m{\mu}
\def\a{\alpha}
\def\b{\beta}

\def\n{\nu}

\def\_#1{_{\scriptscriptstyle #1}}
\def\^#1{^{\scriptscriptstyle #1}}

\def\fpg{4\pi G}

\def\vrf{\varphi}

\begin{document}
\title{Models of modified-inertia formulation of MOND}

\author{Mordehai Milgrom}
\affiliation{Department of Particle Physics and Astrophysics, Weizmann Institute}

\begin{abstract}
Models of ``modified-inertia'' formulation of MOND are described and applied to nonrelativistic many-body systems. Whereas the interbody forces are Newtonian, the expression for their inertia is modified from the Newtonian $m\va$ to comply with the basic tenets of MOND. This results in time-nonlocal equations of motion. Momentum, angular momentum, and energy are (nonlocally) defined for bodies, and the total values are conserved for isolated many-body systems. The models make all the salient MOND predictions. Yet, they differ in important ways from existing ``modified-gravity'' formulations in their second-tier predictions.
Indeed, the heuristic value of the model is in limelighting such possible differences.
The models describe correctly the motion of a composite body in a low-acceleration field even when the internal accelerations of its constituents are high (e.g., a star in a galaxy).
They exhibit a MOND external field effect (EFE) that shows some important differences from what we have come to expect from modified-gravity versions: In one, simple example of the models, what determines the EFE, in the case of a dominant external field, is $\m(\t\av{a\_{ex}}/\az)$, where $\m(x)$ is the MOND ``interpolating function'' that describes rotation curves, compared with $\m(a\_{ex}/\az)$ for presently known modified-gravity formulations. The two main differences are that while $a\_{ex}$ is the momentary value of the external acceleration, $\av{a\_{ex}}$ is a certain time average of it, and that $\t>1$ is an extra factor that depends on the frequency ratio of the external- and internal-field variations. Only ratios of frequencies enter, and $\az$ remains the only new dimensioned constant.
For example, for a system on a circular orbit in a galaxy (such as the vertical dynamics in a disk galaxy), the first difference disappears, since $\av{a\_{ex}}=a\_{ex}$. But the $\t$ factor can appreciably enhance the EFE in quenching MOND effects, over what is deduced in modified gravity. This $\t$ enhancement is important in most applications of the EFE.
Some exact solutions are also described, such as for rotation curves, for an harmonic force, and the general, two-body problem, which in the deep-MOND regime reduces to a single-body problem.

\end{abstract}
\maketitle

\section{Introduction}
From the basic tenets of MOND \cite{milgrom83} follow a number of major predictions concerning aspects of the mass discrepancies in galactic systems, which MOND accounts for without invoking ``dark matter'' \cite{milgrom14a}. Reviews of MOND can be found in Refs. \cite{fm12,milgrom14,milgrom20,mcgaugh20,merritt20,bz22}. However, different formulations that embody these tenets may still differ to varying degrees in making important second-tier predictions. Among the first-tier prediction, one may mention those that pertain to the rotation curves of isolated disk galaxies: asymptotic flatness, the relation between asymptotic speed and total baryonic mass, $M$ (also known as ``the baryonic Tully-Fisher relation''), $V_\infty^4=MG\az$, and the full prediction of rotation curves. Other first-tier predictions are: the mass-velocity-dispersion relation, $MG\az=\eta\s\^4$, with $\eta\sim 1$, in deep-MOND, pressure-supported systems; the presence of effects of an external field on the internal dynamics of a gravitating system -- the external-field effect (hereafter, EFE), etc. Among second-tier predictions, one may count some fine details of the rotation curves (e.g., Refs. \cite{brada95,milgrom12,brown18,pl20,chae22}), the exact value of $\eta$ in the $M~-~\s$ relation, and its dependence on dimensionless attributes of the system; the exact dependence of the effective two-body force on the masses; the exact nature and strength of the EFE, etc.
\par
There are some full-fledged, self-consistent, working formulations of MOND. Among them we have nonrelativistic formulations, such as AQUAL \cite{bm84} and QUMOND \cite{milgrom10}; and we have several relativistic formulations, such as the more recent ones described in Refs. \cite{sz21,milgrom22}. They are all of the type that may be called ``modified gravity'' (MG, hereafter), in that they modify the gravitational (Poisson) action in the nonrelativistic case, or the Einstein-Hilbert action in the case of relativistic versions.\footnote{The definition is not always clear cut. There are theories that can be described equivalently as either a modification of the Einstein-Hilbert action for the metric, with test particles moving on geodesics of the modified metric, or, leaving the gravitational action intact, they modify the particle equation of motion away from the geodesics.}
\par
The presently known nonrelativistic, MG theories (AQUAL and QUMOND) do differ somewhat on second-tier predictions, or else they would not be different theories (e.g., Res. \cite{milgrom10,zf10}).
For example, they exhibit small differences in the EFE (e.g., Refs. \cite{milgrom10,bz15,chaemil22}), small differences in the stability criteria of disks \cite{bmz18}, and small differences in the predicted rotation curves of disk galaxies (e.g. Ref. \cite{chae22}).
However, they do tend to differ only little from each other even on such prediction (less than would be necessary to distinguish between them observationally, at present). For example, Ref. \cite{milgrom14b} derived a {\it general}, deep-MOND, virial theorem for nonrelativistic MG theories, showing that all such theories predict the same value of $\eta$ in the $M-\s$ relation, independent of system parameters, the same dependence of the deep-MOND, two-body force on the two masses, and the same ``$Q$ parameter'' (as defined in Ref. \cite{milgrom12}) for all deep-MOND, disk galaxies.
\par
Concentrating on MG formulations might, however, be too restrictive. For one, because this may lead us to accept the second-tier predictions of such theories as absolute predictions of MOND.
\par
Also, MOND, as we know it now, and as described by any of its presently-known formulations, is, arguably, an effective theory, an approximation, that must have roots in a more fundamental theory -- a ``Fundamond''.
Indeed, beyond their usefulness as practical tools for making predictions and affording self-consistent calculations in MOND, such effective versions of MOND are expected to point the way to a  Fundamond.
\par
It is thus quite valuable to explore a wide variety of models that embody the basic tenets of MOND -- and thus make all its salient prediction -- even if these models are not satisfactory in all regards. Such models can give us some idea of the variety of predictions that MOND can make regarding the second-tier predictions. They can also, potentially, point to more promising paths to the Fundamond.
\par
One possible approach to constructing MOND effective theories is to modify not the gravitational part, but the ``free,'' or ``inertial'' part of the equations of dynamics (gravitational or otherwise).\footnote{Such theories may modify directly the equations of motion, and are not necessarily governed by an action.}
This idea has been discussed at length in the context of MOND \cite{milgrom94,milgrom99,milgrom11}.
\par
In a way of support for the idea, one can recall that physics is replete with examples of acquired, or modified, inertia: inertia of bodies that emerges fully from, or is modified by, the interaction of the body with some omnipresent background medium: the acquisition of particle masses (inertia) by their interaction with the ambient Higgs field in the standard model of particle physics; the renormalization of masses (e.g., in quantum electrodynamics) by interaction with the vacuum; effective modification of the inertia of electrons and holes in  solids; effective inertia of a body moving in a perfect fluid \cite{milgrom06}; etc. etc.
\par
In Ref.  \cite{milgrom99}, I suggested that inertia results from such interaction with the quantum vacuum. The vacuum can, in the least, define an absolute inertial frame, in that bodies can ``know'' that they are accelerated with respect to it, for example via the Unruh effect.
\par
It has to be noted though, that the Einstein-Hilbert action may be considered the free action of the gravitational field. So, an eventual relativistic Fundamond will probably involve modification of all parts of the action.\footnote{This will also be necessary, it appears, if we want to reproduce the observation that gravitational waves follow the same world lines as photons.} Remember, for example, that general relativity modifies all parts of the Newtonian action: On top of the modified choice of gravitational degrees of freedom (a metric instead of a potential), it modifies the Poisson action to the Einstein-Hilbert action, and the Newtonian kinetic Lagrangian, $mv\^2/2$, first to the special relativistic Lagrangian (even in the absence of gravity), $-mc\^2\c(v/c)$,\footnote{Special relativity can be viewed as modifying Newtonian inertia.} and then to the general relativistic one. However, here we consider only the nonrelativistic limit, where the gravitational field may be considered static. I thus restrict myself to modifications of only the kinetic part of the equations of motion.
\par
Following a seed idea in Ref. \cite{milgrom11}, I construct here a class of effective, nonrelativistic models of ``modified inertia'' that embody the basic axioms of MOND. The equations of motion describe a system of many (pointlike) bodies interacting through some (not necessarily gravitational) force. These equations are time nonlocal. (See the discussion in Ref. \cite{milgrom94} as to why such nonlocality arises naturally in modified-inertia theories.)
\par
The models are described in  Sec. \ref{models}, where I also show that the equations of motion obey conservation laws of momentum, energy, and angular momentum, with certain modified definitions of these quantities. Here, I also consider the two-body problem, and scaling laws in the deep-MOND limit.  Further details are worked out in Sec. \ref{construction}. In Sec. \ref{examples}, I discuss some consequences of the models: the prediction of rotation curves of disk galaxies, exact solutions for harmonic forces, the equation of motion of composite bodies, and the MOND external-field effect.
Section \ref{discussion} is a discussion.
\section{Models  \label{models}}
Consider a system of (pointlike) bodies of masses $m\_p$, with the Newtonian force (not necessarily gravitational) of body $q$ on body $p$ being $\vf\_{pq}(\vr\_p,\vr\_q)$, and $\vf\_{qp}=-\vf\_{pq}$.\footnote{We usually deal with forces, whereby $\vf\_{qp}$ is along $\vr\_{pq}=\vr\_p-\vr\_q$, and depends on $|\vr\_{pq}|$.} We want the trajectories of the bodies, $\vr\_p(t)$, to be dictated by MOND. \footnote{I use ``trajectory'' for the path $\vr(t)$, including its time history, and ``orbit'' for the collection of points covered by the trajectory.}
\par
Along its trajectory, body $p$ is subject to the total, time-dependent force
\beq \vF_p(t)=\sum\_{q\not = p}\vf\_{pq}[\vr\_p(t),\vr\_q(t)] .   \eeqno{totalforce}
Newtonian dynamics tell us that the acceleration $\va(t)$ to which the body is subject at time $t$ depends only on the force at time $t$ (for now I omit the subscript $p$) according to $m\va\_N(t)=\vF(t)$, which in Fourier space  reads $m \haNo=\hFo$. Hatted quantities are Fourier components:
\beq \hro=\int\_{-\infty}\^{\infty}\vr(t)e\^{-i\o t}dt,~~~~\vr(t)=\frac{1}{2\pi}\int\_{-\infty}\^{\infty}\hro e\^{i\o t}d\o,   \eeqno{fourier}
with the velocity and acceleration Fourier components $\hvo=i\o\hro,~\hao=-\o\^2\hro$.
\par
We want to modify the Newtonian relation, in accordance with the basic axioms of MOND, by modifying the left-hand, inertia term. We can then write, quite generally,
\beq m\hao\I[\{\hat\vr\},\o,\az]=\hFo,  ~~~~~{\rm or}~~~~~ \hao\I[\{\hat\vr\},\o,\az]=\hat\va\_N(\o),\eeqno{law}
where the inertia functional, $\I[\{\hat\vr\},\o,\az]$, is a dimensionless functional {\it of the whole trajectory} -- designated $\{\hat\vr\}$ to distinguish it from the number $\hro$. $\I$ can also be an explicit function of $\o$, and it is constructed using $\az$ as the only dimensioned constant (in a minimalist view of MOND, we want $\az$ to be the only new, dimensioned constant appearing).
\par
To avoid possible confusion, note that the right-hand side of Eq. (\ref{law}) is not calculable directly from Newtonian dynamics. It is not the Newtonian acceleration on Newtonian orbits; it is rather the Newtonian acceleration on the MOND orbits, which are not known {\it a priori}. Thus, these equations have to be solved self-consistently.
\par
As has been discussed in detail, e.g., in Refs. \cite{milgrom99,milgrom19a}, it is natural in MOND to posit the existence of an absolute inertial frame  with respect to which the absolute accelerations we call $\va(t)$ here, are measured. For example, as argued in Ref. \cite{milgrom99}, the quantum vacuum may define such a frame.
\par
To comply with the MOND basic tenets, $\I[\{\hat\vr\},\o,\az]$ has to approach unity in the formal limit $\az\rar 0$, so as to restore the Newtonian relation when all quantities with dimensions of acceleration are much larger than $\az$. In the gravitational context, at least, MOND is posited to become space-time scale invariance in the opposite limit \cite{milgrom09}, defined by $\az\rar\infty$ and $G\rar 0$, such that $G\az$ remains fixed; i.e., invariant under $(t,\vr)\rar\l(t,\vr)$. The Newtonian gravitational force scales as $\l\^{-2}$, so $\hFo$ whose dimensions are of force times time, scales as $\l\^{-1}$, while $\hao$ (whose dimensions are of velocity) is invariant to scaling.
This means that in this deep-MOND limit, $\I[\{\hat\vr\},\o,\az]$ has to scale as $\l\^{-1}$. But, under scaling, $\I[\{\hat\vr\},\o,\az]\rar\I[\{\hat\vr\},\o,\l\az]$,\footnote{To see this, first apply to $\I$ scaling of the degrees of freedom by $\l$. Then, make a change of time and length unit -- under which $\I$, being dimensionless, does not change -- multiplying them by a factor $\l$. All the degrees of freedom (so $\{\hat\vr\}$ and $\o$) go back to their original numerical values, but the value of $\az$ in the new units is multiplied by $\l$. In other words, under scaling, $\I[\{\hat\vr\},\o,\az]\rar\I[\{\hat\vr\},\o,\l\az]$.\label{tst}} which means that in this limit $\I$ must be of the form
\beq \I[\{\hat\vr\},\o,\az]\rar \A[\{\hat\vr\},\o]/\az, \eeqno{limba}
where $\A$ is a functional of the orbit, and a function of the frequency, with the dimensions of acceleration, and with no dimensioned constant appearing in its construction.
\par
Note that since for gravity $\hFo\propto m$, the theory obeys the universality of free fall (the weak equivalence principle).
\par
If instead of applying Eq. (\ref{law}) to a self-interacting closed system, or to a body moving in some given force field, we dictate a time-dependent force on a body, $\vF(t)$, one may ask whether we should not make sure that the law of motion is causal, in the sense that
$\va(t)$ should depend only on $\vF(t')$ for $t'\le t$. In linear-response systems, where the cause (input) can be clearly separated from the effect (output), such causality is guaranteed by certain analyticity properties in the complex frequency plane, of the analog of $\I[\{\hat\vr\},\o,\az]$ here. Perhaps the models here can be modified to incorporate such a requirement. But to avoid complications, I assume that we are dealing with a closed system, allowed to evolve on its own; so the forces are determined as part of the solution, and are not dictated at will.
\par
In some applications below, I shall specify $\I[\{\hat\vr\},\o,\az]$ to be of the form
\beq \I[\{\hat\vr\},\o,\az]=\m[\A\_1(\o)/\az,\A\_2(\o)/\az,...],\eeqno{mumu}
where $\A\_a(\o)$ -- which may be called ``nonlocal acceleration parameters
 of the trajectory'' -- are  of the type $\A$ described above. In fact, for simplicity of presentation, and also to make more direct comparison with presently known MG theories, I proceed with just one argument in $\m$.
\par
Then, the interpolating function, $\m(x)$, has to approach unity for high arguments $\m(x\rar\infty)\rar 1$, and become linear in its argument, $\Ao/\az$, in the opposite limit.
\subsection{Uniqueness, the ``initial conditions'' problem, and solution by iteration \label{initialvalue}}
The inertia functional, $\I[\{\hat\vr\},\o,\az]$, has to obey some condition, to ensure a certain desired uniqueness requirement of the solutions.
\par
In the MG theory AQUAL, the MOND gravitational potential is determined from the equation
$\div[\m(|\gf|/\az)\gf]=\fpg\r(\vr)$, where $\r$ is the baryonic density. In one-dimensional configurations such as  spherically symmetric systems, this gives an algebraic relation between the Newtonian acceleration, $\gN$, and the MOND acceleration, $g$, at any position: $g\m(g/\az)=\gN$. The Newtonian acceleration field is unique, given $\r$ and spatial boundary conditions. For the MOND field to be unique, it is necessary that $x\m(x)$ be monotonic. Reference \cite{milgrom86} shows that this monotonicity is also a sufficient condition for the AQUAL equation to have a unique solution under analogs of the Neumann or Dirichlet spatial boundary conditions, for an arbitrary density distribution.
\par
There is an analogous requirement of our models here. Consider a physical trajectory, $\vr(t)$, of a body of mass $m$, in a Newtonian force field $\vF(\vr)$.
We do not want another trajectory, $\vr\^\l(t)\equiv\vr(\l t)$, to also be a solution of the equations of motion. Such a solution would have the same orbit as the former, but with the velocities $\vv\^\l(\vr)=\l\vv(\vr)$ and accelerations $\va\^\l(\vr)=\l\^2\va(\vr)$. We would have, for example, in the case of a circular orbit in an axisymmetric potential, more than one orbital velocity at the same radius. This, in turn, will not give a unique prediction for the rotation curve in a disk galaxy.
\par
For the undesired trajectory, $\vr\^\l(t)$, we can write the relevant quantities appearing in the equations of motion in terms of those of $\vr(t)$: For the frequency $\o$ in the spectrum of $\vr\^\l(t)$ we have $\hat\vF^\l(\o)=\l\^{-1}\hat\vF(\o')$ and $\hat\va\^\l(\o)=\l\hat\va(\o')$, where $\o'\equiv\o/\l$ is the matching frequency in the spectrum of $\vr(t)$. Also, on dimensional grounds it can be shown that $\I^\l[\{\vr\^\l\},\o,\az]=\I[\{\vr\},\o',\l\^{-2}\az]$.
Thus, if $\vr\^\l(t)$ satisfies Eq. (\ref{law}), we have
\beq m\l\^2\hat\va(\o')\I[\{\hat\vr\},\o',\l\^{-2}\az]=\hat\vF(\o').  \eeqno{hature}
But, because $\vr(t)$ satisfies the equations of motion, Eq. (\ref{hature})
is also satisfied for all $\o'$ with $\l=1$.
To avoid the above unwanted nonuniqueness it is necessary that, at least for single-frequency orbits,
\beq U(\az,\{\hat\vr\},\o')\equiv\az\^{-1}\I[\{\hat\vr\},\o',\az]  \eeqno{hatul}
be monotonic as a function of $\az$, so the left-hand side of Eq. (\ref{hature}) cannot take the same value for different values of $\l$.\footnote{In the multifrequency case, nonuniqueness can be avoided even if $U$ is not monotonic, because there may not be a single $\l$ for all frequencies.}
In the special case where $\I[\{\hat\vr\},\o,\az]=\m[\A(\o)/\az]$, this necessary uniqueness requirement is tantamount to $x\m(x)$ being monotonic, as in the case of AQUAL.
\par
It is useful to require that $U(\az,\{\hat\vr\},\o)$ is monotonic in $\az$ for all physical trajectories and all frequencies, not only for single-frequency trajectories (see, e.g., an application in Sec. \ref{twobody}).
From the basic tenets of MOND, $U\propto \az\^{-1}$ for $\az\rar 0$, and $U\propto \az\^{-2}$ for $\az\rar\infty$. So $U$ is decreasing with $\az$, and takes up all values between $0$ and $\infty$ for all trajectories and for all frequencies.
\par
Another matter of principle concerns the question of ``initial conditions'':
in Newtonian dynamics, which is local, the theory leads to second-order differential equations, which are solved to propagate the system from some values of the positions and velocities of all constituents at some given time. This is not the situation with time-nonlocal theories. These may be thought of as a sieve -- some conditions on the full history -- that pinpoint the physical system histories out of the many imagined histories. Technically, this is much more difficult to apply, especially given that MOND is also nonlinear even in the nonrelativistic regime. But this does not argue against a theory.
\par
It would be interesting to establish whether in the present models it is still the case that a unique history is defined by dictating the positions and velocities of all particles at some time $t$.
I will show below that this is indeed the case for a particle in a harmonic force field. It is also the case for a body in a constant force field, in which the acceleration is also constant.\footnote{In this case $\vF(\o)\propto \d(\o)$, and so $\va(\o)\propto \d(\o)$.}
But is it the case in general?
\par
We can, in principle, add this as a requirement on $\I$, but we cannot be sure that there is a choice of $\I$ that satisfies it in addition to the other requirements.
\par
We may come closer to an answer by considering a gedanken process of solving the equations of motion by iteration.
For all bodies in the system, take some values of $\vr\_p$ and $\vv\_p$ at some time $t\_0$, which I shall refer to as the ``initial values'', even though in the present framework there is no initiation of the system.
Start with some initial-guess trajectories $\vr\_p(t)$ and $\vv\_p(t)$ that take the initial values at $t\_0$. Substitute these trajectories in the right hand side to calculate the initial guess for $\hat\vF_p(\o)$. Then, in one option, try to calculate the corresponding $\hat\va\_p(\o)$  from the equations.
In another option, substitute the initial guess also in $\I_p$ on the left-hand side, then obtain $\hat\va\_p(\o)$  directly. From $\hat\va\_p(\o)$ calculate $\va\_p(t)$; then, use it to integrate the trajectories forward and backward in time from the initial conditions. These will not generally coincide with our initial guess, but will be our second-iteration trajectories, used for the next iteration.
There are many possible iteration schemes. If there is one for which the process converges to a unique set of trajectories, we have established our goal, since all the trajectories on the iteration path satisfy the initial conditions exactly.
\par
One promising iteration scheme involves also changing the value of $\az$ ``adiabatically'' in the iteration process, as follows: start with the (unique) Newtonian trajectory that satisfies the initial conditions; calculate the nonlocal acceleration for all frequencies; and start with a value of $\az$ that is (much) smaller than all these. The field equations are then satisfied to a desired accuracy by the Newtonian solution. Use this as the initial guess, then increase $\az$ a little and calculate the new trajectories as described above, possibly iterating until convergence for this new value of $\az$ is achieved (which is what I mean by changing $\az$ adiabatically). (In the iteration process we also need to let the frequencies change, since the frequencies of the sought-after solution are, generally, not those of the initial guess.) Then, increase $\az$ more, and repeat, until the actual value of $\az$ is obtained. In this way, we can be sure that the guess is as close as we want to the convergence value.
If this process does converge uniquely, it also establishes an interesting path between the Newtonian and MOND solutions.
\par
We cannot, at present, be sure that any iteration scheme actually converges, that there are no bifurcation points on the iteration path that could lead to multiple convergent solutions, nor that all MOND trajectories can be reached by starting from Newtonian trajectories. But at least this can provide a means to probe the question numerically, and it can also afford solving the equations of motion, at least for simple enough systems, such as few-body problems.
\par
If it can be established that in some version of the model the standard initial conditions determine the solution uniquely, we can say that the theory determines the future of a system from only the initial conditions at some finite time, even though, unlike the standard case, the (putative?) past of the system also enters the dynamics and must be part of the procedure for determining the future.
\subsection{Conservation laws}
We can define a quantity that stands for the momentum, $\vP(t)$, of a body, in such a way that $\frac{d\vP}{dt}=\vF(t)$, or in terms of the Fourier components $i\o\hat\vP(\o)=\hFo$. Thus, from Eq. (\ref{law})
\beq \hat\vP(\o)=m\hat\vv(\o)\I[\{\hat\vr\},\o,\az].  \eeqno{moment}
While $\vP$ depends on time, it is not defined locally by the instantaneous velocity but, rather, depends also on the whole trajectory via $\I$ that appears in its definition.
\par
Because, for an isolated system, $\sum\_p\vF_p=0$, the total momentum in the system is conserved: $d(\sum\_p\vP_p)/dt=0$.
\par
From this follows that the radius $\vR(t)$, whose Fourier transform is
\beq \hat\vR(\o)=(\sum_p m\_p)^{-1}\sum_p m\_p\hat\vr\_p(\o)\I[\{\hat\vr\_p\},\o,\az],  \eeqno{radius}
moves at a constant speed $d^2\vR /dt^2=0$. $\vR$ can be viewed as a (nonlocal) representation of the ``center of mass''.
\par
We can also define a (nonlocal) kinetic energy, $E_k(t)$, by its Fourier transform. It is defined such that
$dE_k/dt=\vv(t)\cdot\vF(t)$. Using the convolution theorem for the Fourier transform of a product
\beq [\a(t)\b(t)]\_\o=\frac{1}{2\pi}\int \a(\o')\b(\o-\o')d\o',  \eeqno{convol}
we have from Eq. (\ref{law})
\beq \hat E_k(\o)=\frac{m}{2\pi}\int \frac{\o'}{\o} \hat\vv(\o-\o')\cdot\hat\vv(\o')\I[\{\hat\vr\},\o',\az]d\o'.  \eeqno{kina}
\par
If the interbody forces are derivable from potentials, $\f\_{pq}(\vr\_{pq})$, with $\vf\_{pq}=-\vf\_{qp}=-d\f\_{pq}(\vr_{pq})/d\vr_{pq}$ ($\vr\_{pq}=\vr\_p-\vr\_q$), we have $\sum\_p\vF_p\cdot\vv\_p=-d\f/dt$, where
$\f(t)=\sum\_{p<q}\f\_{pq}[\vr_{pq}(t)]$, and the total energy $\f+E_k$ is conserved ($E_k=\sum\_p E^k_p$).
\par
Similarly, we define the angular momentum of a body so that its time derivative is $\vr\times\vF$:
\beq \hat \vJ(\o)=\frac{m}{2\pi}\int \frac{\o'}{\o} \hat\vr(\o-\o')\times\hat\vv(\o')\I[\{\hat\vr\},\o',\az]d\o'.  \eeqno{anga}
For a many-body system, the total angular momentum defined in this way is conserved if $\vf\_{pq}$ are along $\vr\_{pq}$.
\par
For high-acceleration trajectories, for which $\I\rar 1$, $\vP(t)$,
$E_k(t)$, and $\vJ(t)$ reduce to the standard expressions, $m\vv(t)$, $(1/2)m\vv\^2(t)$, and $ m\vr(t)\times\vv(t)$, respectively.\footnote{To a body with $\va(t)\equiv 0$ (in the underlying absolute inertial frame), the model assigns vanishing momentum, kinetic energy, and angular momentum. This does not contradict any observations.}
\subsection{Two-body problem \label{twobody}}
Consider an isolated system of two bodies of masses $m\_1$, $m\_2$ and trajectories $\vr\_1(t)$, $\vr\_2(t)$, interacting via a force $\vF(\vr\_{12})$ acting on body 1, along $\vr\_{12}=\vr\_1-\vr\_2$ (the force on $m\_2$ is $-\vF$).
We saw that the radius defined in Eq. (\ref{radius}) moves with a constant speed; so let us work in the Galilei frame where it vanishes and serves as our origin.
From its definition we see that in the two-body system
\beq m\_1\hat\vr\_1(\o)\I[\{\hat\vr\_1\},\o,\az]=-m\_2\hat\vr\_2(\o)\I[\{\hat\vr\_2\},\o,\az].  \eeqno{gushret}
This does not mean, in general, that the origin is always on the line connecting the two masses, since parallelism of the Fourier components does not imply that of the positions themselves.
Are then the two bodies collinear with the origin, and do the distances of the two bodies from the origin always have the same ratio, as in Newtonian dynamics? Try if $\vr\_1(t)=-\a\vr\_2(t)$, with some constant $\a$ [so $\hat\vr\_1(\o)=-\a\hat\vr\_2(\o)$], is consistent with Eq. (\ref{gushret}). Arguments similar to those in Footnote \ref{tst} show that $\I[\{\a\hat\vr\_2\},\o,\az]=\I[\{\hat\vr\_2\},\o,\az/\a]$,
so the question translates to the following: is there a single $\a$ for which
\beq m\_1\a\I[\{\hat\vr\_2\},\o,\az/\a]=m\_2\I[\{\hat\vr\_2\},\o,\az]  \eeqno{gushmush}
for all frequencies?
If $\I$ satisfies our general monotonicity requirement discussed in Sec. \ref{initialvalue}, then there is a unique value of $\a$ for each $\o$.
\par
So, when the orbits are described by a single frequency, such as when the orbits are circular, or when $\vF$ is harmonic, we do have
$\vr\_1(t)=-\a\vr\_2(t)$. When there is more than one frequency involved, there is not, in general, a single value of $\a$ that matches them all (unless $\I$ does not depend explicitly on $\o$).
\par
However, in the deep-MOND regime -- where the general expression (\ref{limba}) applies, with $\Ao\ll\az$ for both bodies -- the general two-body problem simplifies, and -- as in the Newtonian regime -- can be reduced to the problem of a single body in a force field $\vF(\vr\_{12})$, since Eq. (\ref{gushmush}) holds, for all frequencies, with a single $\a=(m\_2/m\_1)\^{1/2}$.
Defining all the relevant quantities for the relative trajectory $\vr\_{12}(t)$, using all our model definitions, we have: $\hat\va\_{12}(\o)=(1+\a^{-1})\hat\va\_1(\o)$, $\A_{12}(\o)=(1+\a^{-1})\A\_1(\o)$. We then find that the relative trajectory satisfies the deep-MOND limit of Eq. (\ref{law})
\beq \bar m\hat\va\_{12}(\o)\frac{\A_{12}(\o)}{\az}=\hFo, \eeqno{lawtwo}
with the reduced mass
\beq  \bar m=\frac{m\_1 m\_2}{(m\_1\^{1/2}+m\_2\^{1/2})^2},  \eeqno{reducedmass}
and $\vF(t)$ is, naturally, defined along the relative trajectory.\footnote{The inverse also holds. Namely, if $\vr\_{12}(t)$ is a solution of Eq. (\ref{lawtwo}) with the reduced mass (\ref{reducedmass}), then $\vr\_2(t)=(1+\a)\^{-1}\vr\_{12}(t)$ and $\vr\_1(t)=-\a(1+\a)\^{-1}\vr\_{12}(t)$ solve the MOND equation with their respective masses.}
\par
So far, we have not specified the dependence of the force on the distance. Applying this result to a self-gravitating binary on a {\it circular} orbit -- which involves a single frequency, and where the magnitude of all quantities is constant -- our models give a velocity difference,
\beq V^4_{12}= (q\_1\^{1/2}+q\_2\^{1/2})^2MG\az,  \eeqno{binar}
where $M$ is the total mass, and $q\_i=m\_i/M$ the mass ratios.
The dependent on the masses is different from what one has in MG theories, which is \cite{milgrom94a,milgrom14b}
\beq V^4_{12}= \frac{4}{9}\frac{(q\_1+q\_2)\^{2}}{q\_1\^2q\_2\^2}[(q\_1+q\_2)\^{3/2}-q\_1\^{3/2}-q\_2\^{3/2}]^2MG\az .  \eeqno{binarmg}
The MOND mass-asymptotic-speed relation (``baryonic Tully-Fisher relation''), $V_{12}^4=MG\az$, is gotten, for both expressions, when one mass is much smaller than the other.
\par
Unlike expression (\ref{binar}), which is valid only for circular orbits, expression (\ref{binarmg}) is valid for an arbitrary orbit (in MG), if $V^2_{12}$ is understood as the mass-weighted mean-squared velocity difference on the trajectory: $V^2_{12}=\av{(\Delta V)^2}$.
But the case of a general trajectory in a gravitationally held binary in our models involves many frequencies and is not obviously solvable analytically (except, perhaps in some epicyclic approximation). But one can argue that on such orbits, the average $V_{12}$ is smaller, with the maximum speed occurring at closest approach and the smallest occurring at the largest separation, with weightier contribution to the average from the latter, where the binary spends more time.
\par
In a central force field, the acceleration (unlike $d\vP/dt$) is not in the radial direction, so Kepler's second law does not hold in general.
This is, similar to the case in special relativity, which might also be seen as modified inertia {\it vis-a-vis} Newtonian dynamics.
\subsection{Deep-MOND limit for self-gravitating systems}
There are general scaling laws obeyed by self-gravitating systems in the deep-MOND limit, which follow from its space-time scale invariance, and dimensional arguments.
It might be instructive to see how these follow in the context of the present models.
\par
Take a many-body system, of masses $m\_p$, and physical trajectories, $\vr\_p(t)$, all satisfying, for all frequencies, the deep-MOND limit of Eq. (\ref{law}) -- namely, with the form (\ref{limba}) of $\I$. Consider
another system, with masses $m\^*\_p=\a m\_p$, and trajectories $\vr\^*\_p(t)=\c\vr\_p(\b t)$, with $\a,~\b,~\c$ some positive constants, such that these are still in the deep-MOND regime.
The quantities appearing in Eqs. (\ref{law}) and (\ref{limba}) calculated for the new trajectories are then $\hat\vF^*(\o)=(\a\^2/\b\c\^2)\hat\vF(\o/\b)$ (for gravity), $\hat\va^*(\o)=(\b\c)\hat\va(\o/\b)$, and $\A^*(\o)=(\b\^2\c)\A(\o/\b)$ [such that we still have $\A^*(\o)\ll\az$]. This history of the new system is also physical if $\a=\b\^2\c\^2$. Velocities scale as $\vv^*(t)=(\b\c)\vv(\b t)$. So this condition underlies the central deep-MOND $M\propto V^4$ relation: the whole family described above has the same value of $V^4/MG\az$. Reference \cite{milgrom14a} explains why this ratio -- which depends only on dimensionless attributes of the system -- is expected to be of order unity and not change appreciably among systems.
Taking $\b\c=\a=1$ corresponds to space-time scaling (in which the masses do not scale).

\section{Construction of the nonlocal acceleration $\A$  \label{construction}}
I now concentrate on the special case where $\I$ is described by Eq. (\ref{mumu}), and consider the construction of the nonlocal acceleration parameters, collectively designated $\Ao$.
We already stated that $\Ao$ has to be of the dimensions of acceleration, and has to be constructed from the orbit (and $\o$) without using dimensioned constants. What other constrains can we cast on its choice?
\subsection{Symmetry requirements}
To ensure translation and Galilei invariance of this model, we construct $\Ao$ from $\hao$, because it is invariant to these [depending only on $\va(t)$]. Formally, under translations by $\vr\_0$, and a Galilei boost by $\vv\_0$, under which $\vr(t)\rar\vr(t)+\vr\_0+\vv\_0 t$, we have $\hao\rar\hao-2\pi (\o\^2\vr\_0-i\o\vv\_0)\d(\o)$, which equals $\hao$ if not applied (as a distribution) to functions that diverge as $1/\o$ or faster, at $\o=0$.
\par
Rotational invariance dictates that only scalars, such as $\hao\cdot\hao$, or $\hao\hat \cdot\va\^*(\o)$ appear.
\par
We also require time-translation invariance of the inertia term, so that trajectories {\it in a time-independent} force field are not affected by a time shift; namely, if $\vr(t)$ is a trajectory, so is $\vr(t+t\_0)$.
Under time translations, $\hao\rar e^{i\o t_0}\hao$. To this, constructs such as $\hat\va(\o\_1)\hat\va(\o\_2)...$ are invariant, provided $\o\_1+\o\_2+...=0$.
For example, $|\hao\cdot\hao|$ and  $\hao\hat \va\^*(\o)$ are invariant. For the sake of concreteness, I shall construct $\Ao$ from $|\hao|$, where the absolute value in $|\hat\va|=(\hat\va\cdot\hat\va\^*)\^{1/2}$ is both in the complex and the vectorial sense, and from  $|\hat\va\cdot\hat\va|\^{1/2}$ (here, the absolute value is in the complex sense), which both satisfy all the above requirements, in addition to invariance to time and space reflections.
\subsection{Explicit dependence on $\o$?  \label{odepend}}
If $\A$ does not depend explicitly on $\o$; i.e., if, for a given trajectory, all frequencies are affected by the same factor $\m(\A/\az)$ -- for example, if $\A\propto\int |\hao|d\o$, the theory can be simply solved for trajectories in a static external field.
Take some Newtonian trajectory $\vr\_N(t)$.
Then, the trajectory $\vr(t)=\vr\_N(\b t)$ is a solution of the field equations for some unique value of $\b>0$. For the above MOND trajectory, $\vF(t)=\vF_N(\b t)$, so $\hFo=\b\^{-1}\hat\vF_N(\o/\b)$. Also, $\hro=\b\^{-1}\hat\vr\_N(\o/\b)$,
$\hao=\b\hat\va\_N(\o/\b)$, and since $\A$ does not depend on $\o$, we have on dimensional grounds
$\A=\b^2\A_N$, where $\A_N$ is the value of $\A$ for the Newtonian trajectory.\footnote{If $\A$ does depend explicitly on $\o$ we have $\Ao=\b^2\A_N(\o/\b)$, and our argument does not go through.} Thus, given that $\vr\_N(t)$ satisfies the field equations with $\m\equiv 1$, $\vr(t)$ satisfies them, provided
\beq  \b\^2\m(\b^2\A_N/\az)=1.   \eeqno{beta}
This equation has a unique solution for $\b>1$. This is because $x\m(x)$ is required to be increasing, and it takes up all values from 0 to $\infty$, as discussed in Sec. \ref{initialvalue}.
\par
This MOND trajectory has the same orbit as the Newtonian one, but with velocities scaled up by a factor $\b$ and accelerations scaled up by a factor $\b\^2=1/\m(\A/\az)$.
\par
The correspondence is one to one: if $\vr(t)$ solves the equations of motion and $\A$ is its frequency-independent, nonlocal acceleration, define $\b$ such that $\b\^2\m(\A/\az)=1$. Then, $\vr\_N(t)=\vr(t/\b)$ solves the Newtonian equations.
\par
Such a theory would satisfy the basic tenets of MOND, and imply its salient predictions.
However, it is too restricted, and, importantly, it does not produce the correct center-of-mass motion of
bodies with high intrinsic accelerations (e.g., stars) in a low-acceleration field (e.g., of a galaxy).
\par
We will thus require some appropriate, explicit dependence of $\A$ on $\o$, as discussed below.
\subsection{Decoupling of the frequencies? \label{decoupling}}
If $\Ao$ is {\it a function} of $\hao$ (and $\o$) -- i.e., depends only on $\hao$ at the same frequency, the theory decouples different frequencies; i.e., it modifies the acceleration frequency by frequency. Such a theory would satisfy the basic MOND axioms, and predict the salient MOND predictions; but it would practically exhibit no MOND EFE (see Sec. \ref{efe} below for more details). The EFE would act only if the frequency of the external force were the same as that of the internal one, which is hardly ever the case. This demonstrates that an EFE can, in practice, be avoided in a theory satisfying the basic tenets of MOND. But there are strong indications that an EFE is required in MOND (e.g., to account for the fact that no MOND effects have shown up in terrestrial experiments). In the framework of the present model, this requires that different frequencies (e.g., of the external and internal accelerations) are coupled in $\Ao$.
\subsection{Heuristic example of a choice of $\Ao$}
A way to define $\Ao$ under all the above requirements, using only $|\hat
\va(\o')|$ for concreteness (see below on the effects of using $|\hat\va\cdot\hat\va|\^{1/2}$ as well) is
\beq \Ao=\frac{1}{2\^{3/2}\pi}\int\_{-\infty}\^{\infty}\t\left(\frac{\o'}{\o}\right)|\hat
\va(\o')|d\o'=\frac{1}{\sqrt{2}\pi}\int_{0}^{\infty}\t\left(\frac{\o'}{\o}\right)|\hat
\va(\o')|d\o'=\frac{1}{\sqrt{2}\pi}\int_{0}^{\infty}\t\left(\frac{\o'}{\o}\right)|\hat
\vr(\o')|{\o'}\^2d\o', \eeqno{v}
where $\t(y)$ is a dimensionless function, which I take to be symmetric, $\t(-x)=\t(x)$. In the second equality in Eq. (\ref{v}), I use the fact that $|\hao|$ is symmetric, since $\va(t)$ is real, and hence, $\hat \va^*(\o)=\hat \va(-\o)$.\footnote{$|\hao\cdot\hao|\^{1/2}$, which can also be used in the construction of $\Ao$, is also symmetric under $\o\rar -\o$.}
\par
The normalization of $\t(y)$ is degenerate with that of $\az$. We pick the normalization such that $\t(1)=1$ because then, as we shall see below, the standard value of $\az$ and the form of $\m(x)$ that have been routinely used in rotation-curve analysis, apply.

\section{Examples  \label{examples}}
As a general comment, we note that it is impractical, in general, to solve a many-body system in full, but as is usually realized, there are problems where we can apply a mean-field approximation, where we consider the motion of a single body (e.g., a star in a galaxy) in the combined mean field of all the others assumed time independent.\footnote{If the (Newtonian) force field in which the body is moving is derivable from a potential, $\vF(t)=-\gf[\vr(t)]$, with the above definition of the kinetic energy, we have conservation of the energy $d(E_k+\f)/dt=0$ along a trajectory.}
\par
It may go without saying, but I emphasize that such a mean field does not define an effective, MOND force field -- or an acceleration field in the case of gravity -- as is the case of MG. This is because the accelerations of different bodies, at the same position, may differ from each other, depending on their kinematics, and, in the present, time-nonlocal case, on details of their trajectories. What is common to all bodies at the same position, $\vr$, is their $d\vP/dt=\vF(\vr)$ in a given force field, or $m\^{-1}d\vP/dt=\va\_N(\vr)$, in a Newtonian, acceleration field $\va\_N(\vr)$.\footnote{In special relativity, where the equation of motion is $d\vp/dt=md(\c\vv)/dt=m\c[\va+\c\^2c\^{-2}(\vv\cdot\va)\vv]=\vF$, it is also the case that an electric field does not define an acceleration field for an electron, say. What is common to all electrons at the same position is $d\vP/dt$.}
\par
In some important applications, the motion of the constituents is described, to a good approximation, as the combination of several trajectories, with distinct frequencies, or of bunches of frequencies (which we lump into one for simplicity),
\beq \vr(t)=\sum_k \vr\_k(t), ~~~~~~~~ \vr\_k(t)=\frac{1}{\sqrt{2}}(\bar\vr\_k e\^{i\o\_k t}+\bar\vr\^*\_k e\^{-i\o\_k t}), \eeqno{mufre}
where $\bar\vr\_k $ are complex amplitudes, normalized such that for a circular orbit, the orbital radius is $|\bar\vr\_k|=(\bar\vr\_k\cdot\bar\vr\^*\_k)\^{1/2}$. Furthermore, in such instances, the (Newtonian) force on a constituent separates, to a good approximation, into forces that depend on the separate component trajectories:
\beq \vF[\vr(t)]\approx \sum_k\vF_k[\vr\_k(t)].   \eeqno{separa}
Consider, e.g., a small many-body system, such as a binary star, or a star cluster, moving on a ``large'' orbit, of scales much larger than its size, in the field of a large body, such as a galaxy. Then, one component of the motion of a constituent corresponds to the large motion, and the force associated with it is that applied by the large body, while the other components describe the motion of the constituent within the body, and is dictated mainly by the inter body force. So, there is a natural separation of trajectories and the associated forces.
\par
Another example is the vertical dynamics in a thin galactic disk. Each constituent -- such as a star, or a gas element, may be viewed as performing a combination of two motions: one in the mean field of the galaxy characterized by an orbital size close to the galactic radius, $R$, at the position of the body and the other, a vertical motion, governed by the local density distribution vertical to the disk. Provided that we are probing heights much smaller than $R$,  and that the characteristic frequencies are distinct, the separation of forces, as above, is a good approximation. A separation may also be justified if we are considering separately motions along different axes, i.e., if the vectorial components of Eq. (\ref{law}) are described by different frequencies (e.g., in a triaxial harmonic field).
\par
We can then write separate equations for the different components that are coupled though $\A(\o\_k)$, which depends on the motions in all the components.
The equations of motion (\ref{law}) then read

\beq m\hat\va\_k(\o\_k)\m[\A(\o\_k)/\az]=\hat\vF_k(\o\_k).\eeqno{lawla}
Each frequency component describes an elliptical orbit, with
\beq \vr\^2\_k(t)=\bar\vr\_k\cdot\bar\vr\^*\_k+{\rm Re}(\bar\vr\_k\cdot\bar\vr\_k e\^{2i\o\_k t})=\bar\vr\_k\cdot\bar\vr\^*\_k+|\bar\vr\_k\cdot\bar\vr\_k|\cos{(2\o\_k t+\vrf\_k)}.  \eeqno{ellap}
So, $|\bar\vr\_k|$ is the root-mean-square radius, and $|\bar\vr\_k\cdot\bar\vr\_k|\^{1/2}$ measures the ellipticity: it vanishes for a circular orbit, and equals $|\bar\vr\_k|$ for a radial orbit.
Then,
\beq \hao=-\sqrt{2}\pi\sum_k \o\_k\^2[\bar\vr\_k\d(\o-\o\_k)+\bar\vr\^*\_k\d(\o+\o\_k)].   \eeqno{jamula}
For $\o>0$, needed in Eq. (\ref{v}), we have (remembering that the product of delta functions with different frequencies vanishes)
\beq |\hao|=\sqrt{2}\pi\sum_k \o\_k\^2|\bar\vr\_k|\d(\o-\o\_k),~~~~~~~|\hao\cdot\hao|\^{1/2}=\sqrt{2}\pi\sum_k \o\_k\^2|\bar\vr\_k\cdot\bar\vr\_k|\^{1/2}\d(\o-\o\_k).   \eeqno{jama}
The two are equal for radial trajectories, and the second vanishes for circular ones.
Proceeding, using only the first for concreteness sake, we have
\beq \Ao=\sum_k \o\_k\^2|\bar\vr\_k|\t\left(\frac{\o\_k}{\o}\right). \eeqno{shiupa}

\par
To determine the motion, we need to know $\Ao$ at the frequencies underlying each component. For the $n$th component,
\beq \A(\o\_n)=\o\_n\^2|\bar\vr\_n|+\sum_{k\not =n} \o\_k\^2|\bar\vr\_k|\t\left(\frac{\o\_k}{\o\_n}\right). \eeqno{shiluta}
Thus, importantly, the acceleration measure $\A(\o\_n)$, which enters that MOND ``magnification factor'' for frequency $\o\_n$, namely, $1/\m[\A(\o\_n)/\az]$, does not equal the acceleration at $\o\_n$ itself, which is $\o\_n\^2|\bar\vr\_n|$, but it picks up contributions from all other frequencies, weighted by $\t(\o\_k/\o\_n)$.

\subsection{Circular orbits -- rotation curves}
For the important example of circular orbits in an axisymmetric field, underlying rotation curves of disk galaxies, there is only one frequency,
and the basic equation (\ref{law})
gives directly the MOND algebraic relation between the Newtonian and MOND accelerations, radius by radius,
\beq a\m(a/\az)=a\_N,  \eeqno{mdar}
where $a\_N$ is the Newtonian acceleration at the orbital radius, $r$, and $a=\o\^2(r) r=V^2(r)/r$. This is as expected from the general theorem for modified-inertia theories \cite{milgrom94}.
Here it follows simply from the general equations of motion (\ref{law}), since circular trajectories (with constant speeds) are characterized only by their radius, $r$, and frequency, $\o$, so for these, $\I$ must reduce to some $\m(\o\^2r/\az)$.
\par
When the models are applied to the case where the forces are gravitational, this describes correctly the rotation curves of galaxies, with $a\_N=g\_N$, the Newtonian gravitational acceleration at radius $r$.

\subsection{Harmonic force}
Another example that is easy to solve is that of the motion of a particle in a harmonic force field, for example, a star moving in the gravitational field of a spherical mass of constant density.
\par
Start with the isotropic case, where $\vF(t)=-k\vr(t)$. In this case,  Eq. (\ref{law}) reads
\beq  \o\^2\m[\Ao/\az]=\o\_0\^2, \eeqno{harmonic}
where $\o\_0=(k/m)\^{1/2}$ is the Newtonian frequency of a particle of mass $m$ in this field.
There are then single-frequency, oscillatory solutions, on elliptical orbits, of the type (\ref{mufre}), with arbitrary values of the six components of the complex amplitude $\vr\_0$, with frequency $\o$ given by
\beq  \o\^2\m(\o\^2 |\vr\_0|/\az)=\o\_0\^2,  \eeqno{osca}
for which there is a unique solution for $\pm|\o|$ for any $|\vr\_0|$. This is, again, because $x\m(x)$ is monotonically increasing from $0$ to $\infty$, as discussed in Sec. \ref{initialvalue}.
We see that the oscillation frequency depends on the root-mean-square radius of the orbit, $|\vr\_0|$, but not on the ellipticity [see Eq. (\ref{ellap})]. Had we also used $|\hat\va\cdot\hat\va|\^{1/2}$ in the construction of $\Ao$, there would also be a dependence on the ellipticity through $|\vr\_0\cdot\vr\_0|$.
\par
As in Newtonian dynamics, given a position $\vr(0)$, and velocity, $\vv(0)$ at some time, say $t=0$, there is a unique solution with these ``initial conditions', since they determine $\vr\_0$: $Re(\vr\_0)=\vr(0)$, and $Im(\vr\_0)=-\o\^{-1}\vv(0)$.
\par
The motion of a particle of mass $m$ in an anisotropic harmonic field can also be easily solved. If the principal force constants are
$k\_n$, and the Newtonian frequencies $\bar\o\_n=(k\_n/m)\^{1/2}$, then the motion is still a combination of harmonic motions in the principal axes $x\_n(t)=\sqrt{2}x\^0\_n\cos(\o\_n t+\vrf\_n)$, with arbitrary amplitudes, $x\^0\_n$,
and phases $\vrf\_n$, but the frequencies are determined from coupled equations
\beq \o\_n\^2\m\_n=\bar\o\^2\_n;~~~~~~~~~~\m\_n=\m\{[x\_n\^0\o\_n\^2+\sum\_{l\not =n} \t\left(\frac{\o\_l}{\o\_n}\right)x\_l\^0\o\_l\^2]/\az\}.  \eeqno{harmoan}
Note that this expression does not tend exactly to relation (\ref{osca}) when all the frequencies are equal. This has to do with our working with sharp delta functions in the Fourier transform.

\subsection{Motion of a composite system in an external field \label{com}}
Consider now the so-called ``center-of-mass-motion problem''. It concerns the fact that objects -- such as stars and gas clouds -- move in a galaxy according to MOND, while their constituents (e.g., ions) are subject to very high accelerations. The problem was solved early on for modified-gravity MOND \cite{bm84}. In the context of the models discussed here, what matters is how the high-acceleration, high-frequency components of the constituent motions affect the low-acceleration, low-frequency components of motion in the galaxy. Take $\o\_1$ to be that of the dominant acceleration in the galaxy and $\o\_2$ to represent the frequencies of the internal motions.
\par
We need the MOND acceleration at $\o\_1$, which is given in our models by $\va(\o\_1)\m[\A(\o\_1)/\az]=\va\_N(\o\_1)$,
where
\beq \A(\o\_1)=\o\_1\^2|\bar\vr\_1|+\o\_2\^2|\bar\vr\_2|\t\left(\frac{\o\_2}{\o\_1}\right). \eeqno{shimasa}
While for the problem at hand , $\o\_1\^2|\bar\vr\_1|\lll\o\_2\^2|\bar\vr\_2|$ (for the sun in the Galaxy, the ratio is $\sim 10^{12}$), we also have
$\o\_1\lll\o\_2$ (also with a ratio of $\sim 10^{12}$). So, if $\t(y)$ decreases fast enough (e.g., exponentially, or even as a power larger than 1), the second term in Eq.(\ref{shimasa}) can be neglected, and we get the desired center-of-mass motion.\footnote{The $\o\_1$ component affects negligibly the $\o\_2$ motion, since $\A(\o\_2)=\o\_2\^2|\bar\vr\_2|+\o\_1\^2|\bar\vr\_1|\t(0)\approx\o\_2\^2|\bar\vr\_2|$.}
\par
Such large frequency ratios are the rule in such circumstances: we are dealing with systems small compared with their orbit size, $|\bar\vr\_2|\ll|\bar\vr\_1|$, but much larger internal accelerations, $\o\_1\^2|\bar\vr\_1|\lll\o\_2\^2|\bar\vr\_2|$; so, clearly, $\o\_1\lll\o\_2$.

\subsection{External-field effect \label{efe}}
The MOND EFE pertains to a subsystem that is falling in the field of some larger mother system. Examples are a dwarf satellite, a globular cluster, or a binary star in the field of a galaxy; a galaxy in a field of a galaxy cluster; or the dynamics perpendicular to a galactic disk, while the region is being accelerated on its circular orbit in the field of the galaxy. MOND predicts, quite generically,\footnote{See Ref. \cite{milgrom14a} for a discussion of the extent to which the EFE follows from the basic tenets of MOND.} that the internal dynamics of the subsystem, with a characteristic acceleration $\va\_{in}$, is affected by the external field, of characteristic acceleration $\va\_{ex}$. This is unlike Newtonian dynamics (and general relativity) where, as long as the external field may be considered constant across the subsystem (so that there are no tidal effects), the internal dynamics is oblivious to the external field in which the subsystem falls freely. This, in general relativity, is an expression of the strong equivalence principle, which is not obeyed in MOND.
\par
We already saw in Sec. \ref{decoupling} that in the present models it is possible to avoid an EFE by adopting $\Ao$ that decouples the different frequencies. But this is undesirable.
\par
The EFE was, in fact, found to be relevant, in varying degrees, in many astrophysical circumstances.
Reference \cite{mm13} finds signs for an EFE in the internal dynamics of some satellites of the Andromeda galaxy.
References \cite{muller19,haghi19} discuss the importance of the EFE in ultradiffuse galaxies.
Also, Refs. \cite{haghi16,hees16,wu15,chae20,chae21} show that the MOND predictions of rotation curved perform somewhat better if one takes into account an EFE from large-scale structure. Reference \cite{asencio22} demonstrates that the Fornax cluster EFE on dwarf-galaxy members makes them vulnerable to tidal distortion and disruption, as observed, and contrary to the expected protection endowed by their putative dark-mater halos. The EFE is also responsible for quenching, practically completely, all MOND effects in experiments on Earth.
\par
The EFE is expected to depend on the relative strength of $\va\_{in}$ and $\va\_{ex}$ and on their values relative to $\az$, and also on their directions and the general geometry of the system. But in the limit where $|\va\_{in}|\ll|\va\_{ex}|$, the presently known modified-gravity versions of MOND -- which are all underlain by an interpolating function $\m(x)$ -- predict that the internal dynamics is essentially Newtonian, with an increased gravitational constant $\bar G\approx G/\m(|\va\_{ex}|/\az)$ (and some anisotropy introduced by the direction of $\va\_{ex}$).
And, importantly, in such theories, the EFE depends only on the momentary value of $\va\_{ex}$ at the position of the subsystem.
\par
In modified-inertia versions, as exemplified by the present models, the EFE acts differently. The two main differences being as follows: 1. It is expected in such time-nonlocal theories that the EFE will depend not on the momentary value of the external field, but, in some way on the full trajectory of the subsystem. 2. Such formulation may introduce dependence of the EFE on other parameters. In the present heuristic models, the EFE depends also on the frequency ratios of the internal and external motions (as we saw is the case also in connection with the center-of-mass motion of the subsystem).
\par
To demonstrate this with a simple example, consider again the two-frequency case described by Eq. (\ref{shimasa}). But now, $\o\_1$ characterizes the intrinsic motions, the dynamics of which we want to describe, while $\o\_2$ characterizes the time variation of the external field. So, we write
\beq \A(\o\_{in})=\o\_{in}\^2|\bar\vr\_{in}|+\o\_{ex}\^2|\bar\vr\_{ex}|\t\left(\frac{\o\_{ex}}{\o\_{in}}\right). \eeqno{shitaba}
\par
In practically all applications we have $\o\_{ex}<\o\_{in}$. In some dwarf satellites of the Milky Way and Andromeda (and presumably other mother galaxies), we estimate $\o\_{ex}\sim\o\_{in}$.
\par
We saw above that to predict the correct center-of-mass motion of composite systems, we need $\A$ to depend on $\o$, as modeled here by $\t(y)\not \equiv 1$. Furthermore, we need $\t(y)$ to be decreasing relatively fast for $y\gg 1$. If, in fact, $\t$ decreases everywhere, including for $y<1$, which is reasonable, we will have $\t(\o\_{ex}/\o\_{in})>1$ [since $\t(1)=1$]. Instead of the external acceleration $\o\_{ex}\^2|\bar\vr\_{ex}|$ deciding the effect, it is the larger $\o\_{ex}\^2|\bar\vr\_{ex}|\t(\o\_{ex}/\o\_{in})$. For the prevalent occurrence of $\o\_{ex}\ll\o\_{in}$ this is $\approx\o\_{ex}\^2|\bar\vr\_{ex}|\t(0)$.
\par
We have no knowledge of the form of $\t(y)$, but unless it behaves unusually below $y=1$, we can expect $\t(0)$ to be of the order of a few.
For example, for $\t(y)=2/(1+y\^2)$, $\t(0)=2$; for $\t(y)= e\^{(1-x)}$, $\t(0)=e$; more generally, for $\t(y)= e\^{(1-x)/q}$, $\t(0)=e\^{1/q}$; etc.
\par
The EFE is simply accounted for in the present models, when the subsystem is moving on a circular orbit, say in a disk galaxy. Then the external field is indeed characterized by a single orbital frequency $\o\_{ex}$, and $\o\_{ex}\^2|\bar\vr\_{ex}|$ is the observed orbital acceleration $a\_{ex}=V^2/r$.
Inasmuch as the internal accelerations at all frequencies are appreciably smaller than the orbital one, and that the internal frequencies are appreciably smaller than $\o\_{ex}$, the models give for all internal frequencies
\beq \hat\va(\o\_{in})\m[\t(0)a\_{ex}/\az]=\hat\va\_N(\o\_{in}). \eeqno{exasa}
Since $\m$ here is a constant (at a given radius in the disk galaxy), the intrinsic accelerations are what one would calculate in Newtonian dynamics {\it for the observed mass distribution and particle positions},\footnote{We reiterate that the right-hand side employs the calculated Newtonian accelerations on the {\it observed} (MOND) orbits, not on what we would calculate as Newtonian orbits.} all multiplied by the approximately constant factor $1/\m[\t(0)a\_{ex}/\az]$. Remember that with the normalization $\t(1)=1$, it is $\m(a\_{ex}/\az)$ that enters the rotation-curve analysis, such that $1/\m(a\_{ex}/\az)$ is the mass discrepancy indicated by the rotation curve at position $r$. So, importantly, it is not this factor that enters the EFE, as is the case in the presently used modified-gravity theories.
\par
For example, for the description of vertical dynamics, or that of wide binaries in the solar neighborhood we have $a\_{ex}/\az\approx 2$, and even a value of $\t(0)$ of a few can have a large impact on $1-\m$,  which determines the departure of the internal dynamics from Newtonian, since $1-\m[2\t(0)]$ can be rather smaller than $1-\m(2)$.
\par
Regarding the vertical dynamics in thin disk galaxies, I already noted at the beginning of the section that one cannot describe such dynamics in terms of some vertical acceleration field. It is well known that different populations of stars, and the gas in the disk,
have different kinematics: they have different velocity dispersions, different scale heights, and hence different characteristic vertical motions, with different characteristic frequencies. Some of these frequencies can be below and some can be above the local orbital frequency around the galaxy. A proper treatment of the vertical dynamics in the framework of the present models must take all this into account.

\section{Conclusions \label{discussion}}
Here, I am propounding a class of models that embody the basic axioms of MOND but, unlike presently known MOND theories, are based on modifying the kinetic, inertial response of bodies to applied forces. These models have a relatively simple, quasialgebraic formulation in terms of the Fourier components of the body's trajectory. This choice was inspired by the observation, made in Ref. \cite{milgrom94}, that such modified-inertia formulations probably have to be time nonlocal. This conclusion was reached on formal grounds. But, also from the physics point of view, such modifications envisage inertia as an acquired attribute, which results from the interaction of the body with some ambient medium, which resists acceleration. A time-local inertia results if the medium cures the effects of the bodies acceleration on timescales much shorter that characteristic kinematic times of the body. But in general, we expect nonlocal effects to result. The model is also inspired by linear-response systems (e.g., in electronics) which are described by an algebraic relation in frequency space between the input, output, and the response function of the device and produce time nonlocality. But, here, in addition to the nonlocality, the equations of motion are nonlinear, so as to be compatible with the axioms of MOND.
\par
I call this construction ``a class of models'', because, as we saw, even within the general form of the equations of motion (\ref{law}) there may be other possible choices of $\A$ as a function of $\o$, and a functional of the trajectory. It may also be possible to construct such models with $\m$ as a function of several variables $\A_a$. For example, we can use $|\hao\cdot\hao|\^{1/2}$ to form a second nonlocal acceleration measure, as in Eq. (\ref{v}), even with its own $\t$ function. Since this quantity vanishes for circular orbits, using it as a second variable, $x\_2$ in $\m(x\_1,x\_2)$ would mean that $\m(x,0)$ is what enters the predictions of rotation curves.
\par
Also, the equations of motion (\ref{law}) can be generalized in various ways. For example, we may generalize $\I$ not to be a number that multiplies $\hao$ but as an operator that acts on it, replacing Eq. (\ref{law}), for example,  by
\beq \int \I[\{\hat\vr\},\o,\o',\az]\hat\va(\o')d\o'=\hFo.   \eeqno{jajaja}
\par
While the construction of the models is somewhat arbitrary at present, these models are helpful especially in pointing to some important possible differences between the second-tier predictions of such theories from those of MG. At present, I see this has their main value. But perhaps, with some deliberation, they will show a (different) way to look for a more fundamental theory for MOND.
\par
We have known that modified-inertia and MG theories make slightly different predictions of the exact rotation curves of galaxies, in that, unlike the latter, the former predict an algebraic relation between the Newtonian and the MOND acceleration at every radius. This is confirmed in the present class of models as seen in Eq. (\ref{mdar}).
We also saw that in the present models, the strength of the EFE, and its mode of action in general, can differ materially in the two classes of theories.
In another example, we saw that the exact dependence on the masses of the velocity difference in a deep-MOND binary on a circular orbit differ between the two classes; the difference is of order unity, but might be detectable.
\par
The first full-fledged, NR, MG formulation of MOND, AQUAL, is highly nonlinear, and harder to solve. It generally requires numerical solution, which is done by iterating from some initial guess, as suggested in Sec. \ref{initialvalue} for the present models.
In contradistinction, QUMOND is ``quasilinear'' in that its solution requires solving only linear differential equations, with one nonlinear, algebraic step in between. In some rough sense, it achieves this by inverting the AQUAL equation from the schematic $\vg\m(|\vg|/\az)=\vg\_N$ to the schematic $\vg=\vg\_N\n(|\vg\_N|/\az)$. Might we also construct modified-inertia models that invert the field equations (\ref{law}), by which
$\hao$ is more easily gotten from the Newtonian trajectories, e.g., a relation of the type
\beq \hao=\haNo\J[\{\hat\vr\_N\},\o,\az].  \eeqno{hututu}
If feasible, this will make life much easier. At present, I see two obstacles to constructing such a formulation: (i) I do not see a way to define conserved momentum, energy, and angular momentum in such a formulation, which I deem necessary for a healthy theory. (ii) In any event, the right-hand, ``Newtonian'' side cannot be calculated as part of Newtonian dynamics, since it has to be calculated for the MOND trajectories, which are not known {\it a priori}. But it is worthwhile to look for a formulation in this vein.

\end{document}